\documentclass[11pt,a4paper]{amsart}

\usepackage{amsmath,amsfonts,amssymb}
\usepackage{mathrsfs}
\usepackage{dsfont}

\usepackage{cases}

\usepackage{svg}
\usepackage{amsthm}
\usepackage{braket}
\theoremstyle{definition}
\newtheorem*{rem}{Remark}

\newtheorem{st}{Proposition}
\newtheorem*{pr}{Proof}
\newtheorem{df}{Definition}

\newtheorem{cor}{Corollary}

\begin{document}

\title{Cohomological beta function}

\author[O. Gamayun]{Oleksandr Gamayun}
\address{Oleksandr Gamayun:\ London Institute for Mathematical Sciences, Royal Institution, 21 Albemarle St, London W1S 4BS, UK}
\email{og@lims.ac.uk}

\author[M. Gritskov]{Maxim Gritskov}
\address{Maxim Gritskov:\ Saint Petersburg State University, Universitetskaya nab. 7/9, 199034 St. Petersburg, Russia}
\email{maximgritskovvl@gmail.com}

\author[A. Losev]{Andrey Losev}
\address{Andrey Losev:\ Shanghai Institute for Mathematics and Interdisciplinary Sciences,
Block A, International Innovation Plaza, No. 657 Songhu Road, Yangpu District, 200433, Shanghai, China}
\email{aslosev2@yandex.ru}

\begin{abstract}
	We propose a cohomological approach to computing the conformal anomaly. Using the example of $J\bar{J}$-deformations of two-dimensional conformal field theories, we reproduce the well-known Cardy formula for the leading contribution to the perturbative beta function as the coefficient of the cocycle that realizes the obstruction to deforming the conformal algebra module structure on the state space. In addition to offering a novel conceptual perspective on the conformal anomaly, the proposed approach is anticipated to provide an efficient tool for computing higher-order coefficients of perturbative beta functions.
\end{abstract}

\maketitle

\date{\today}

\tableofcontents
\section{Introduction}
\subsection{Motivation}
A natural approach to studying deformations of various mathematical structures relies on associating them with a cochain complex. The deformation theory of complexes is well established: infinitesimal deformations correspond to the first cohomology of the adjoint complex, while obstructions manifest as multilinear operations on the first cohomology with values in the second. These operations endow the cohomology of the adjoint complex with the structure of an $L_{\infty}$-\emph{algebra}.

In this work, to a two-dimensional conformal field theory we associate the Chevalley-Eilenberg complex of two Virasoro algebras acting on the space of states. Deformations of the conformal theory triggered by a marginal observable are mapped to certain classes in the first Chevalley-Eilenberg cohomology. In quantum field theory, it is well known that the obstructions to deforming a conformal theory are encoded in the coefficients of the \emph{beta function}. Thus, the beta function can be interpreted in a purely algebraic manner as a set of obstructions to deforming the representation of the conformal algebra. Remarkably, it turns out that constructing the cohomological conformal anomaly requires no local data of quantum field theory, such as correlation functions or ultraviolet cutoffs.

We implement this program for $J\bar{J}$-\emph{deformations} of two-dimensional conformal field theories and demonstrate that the first obstruction to deforming a conformal algebra module indeed coincides with the well-known \emph{Cardy formula} for the quadratic contribution to the beta function.
\subsection{Deformation theory of Virasoro modules}
Consider a $d=2$ conformal field theory with the state space $\mathcal{H}$, which forms a module over the \emph{conformal algebra} $\mathfrak{Conf}_{2}=\mathfrak{Vir}\oplus\overline{\mathfrak{Vir}}$. This structure is defined by the action of the generators $L_{m}, \bar{L}_{m}$ and the central elements $C, \bar{C}$ on $\mathcal{H}$, satisfying the commutation relations:
\begin{equation}
\begin{aligned}
    [L_{m},L_{n}]=(m-n)L_{m+n}+\delta_{m+n,0}\cdot\frac{m^{3}-m}{12}\cdot C\,,\\
    [\bar{L}_{m},\bar{L}_{n}]=(m-n)\bar{L}_{m+n}+\delta_{m+n,0}\cdot\frac{m^{3}-m}{12}\cdot \bar{C}\,.
\end{aligned}
\end{equation}

Under a first-order marginal deformation, the CFT remains conformal; this induces a \emph{deformation of the conformal algebra representation} on $\mathcal{H}$. Such deformations and their obstructions are described using methods of homological algebra.
\begin{df}
    The Chevalley-Eilenberg complex of the algebra $\mathfrak{Conf}_{2}$ with coefficients in $\mathcal{H}$ is the $\mathbb{Z}$-graded vector space \cite{chevalley1948cohomology}
    \begin{equation}
    \label{Vir_complex}
    \begin{aligned}
    \mathrm{CE}^{\bullet}_{\mathrm{Q}}(\mathfrak{Conf}_{2},\mathcal{H})=\mathrm{Hom}(\mathrm{S}^{\bullet}\mathfrak{Conf}_{2}[1],\mathcal{H})\,,
    \end{aligned}
    \end{equation}
    equipped with the differential $\mathrm{Q}= \mathcal{Q} + \bar{\mathcal{Q}}$, where the chiral part is defined as
    \begin{equation}
    \label{chiral_CE}
    \begin{aligned}
    \mathcal{Q}=\gamma \,C+\sum_{m}c^{\,m}L_{m}-\frac{1}{2}\sum_{m,n}(m-n)c^{\,m}c^{\,n}\frac{\partial}{\partial c^{\,m+n}}-\\-\sum_{m}\frac{(m^3-m)}{24}c^{\,m}c^{\,-m}\frac{\partial}{\partial \gamma}\,.
    \end{aligned}
    \end{equation}
    Here, $c^{\,m}$ are the odd dual basis elements corresponding to $L_{m}$, while $\gamma$ is the ghost variable dual to the central charge $C$. The antichiral part of the differential is defined completely analogously and contains the ghosts $\bar{c}^{\,m}$ dual to $\bar{L}_{m}$, as well as the ghost $\bar{\gamma}$ dual to the antichiral central charge $\bar{C}$.
\end{df}
\begin{rem}
    The space $\mathrm{S}^{\bullet}(\mathfrak{Conf}_{2}[1])$ consists of polynomials in anticommuting variables corresponding to the generators of the conformal algebra. Accordingly, the Chevalley-Eilenberg complex consists of linear maps from this polynomial space to $\mathcal{H}$. For example, if $\mathcal{H}=\mathbb{C}$ with the trivial action, an element of the Chevalley-Eilenberg complex is
    \begin{equation}
        \sum_{m}\frac{m^{3}-m}{24}c^{\,m}c^{\,-m} \in \mathrm{CE}^{2}_{\mathrm{Q}}(\mathfrak{Conf}_{2},\mathbb{C})\,.
    \end{equation}
    This is the familiar expression for the Gelfand-Fuchs cocycle \cite{GelfandFuks1968} defining the Virasoro central extension of the Witt algebra. Viewed as a cochain on the conformal algebra $\mathfrak{Conf}_2$, this cocycle is exact and equals $-\mathrm Q\gamma$. Although the sum is infinite, only finitely many terms contribute when it is evaluated on any polynomial in $\mathrm{S}^{\bullet}(\mathfrak{Conf}_{2}[1])$.
\end{rem}
\begin{rem}
    Generally speaking, the central charges $C$ and $\bar{C}$ are independent and for our purposes are not required to be equal.
\end{rem}
\begin{df}
    Along with the complex \eqref{Vir_complex}, we will introduce the Chevalley-Eilenberg complex with values in the representation $\mathrm{End}(\mathcal{H})$, whose differential is defined as $\{\mathcal{\mathrm{Q}},\cdot\}$. Here, $\{\cdot,\cdot\}$ denotes the graded commutator of operators on the Chevalley-Eilenberg complex:
    \begin{equation}
    \{A,B\}=AB-(-1)^{|A||B|}BA\,,
    \end{equation}
    where $|A|$ and $|B|$ are the degrees of the $\mathrm{End}(\mathcal{H})$-valued cochains $A$ and $B$. In particular, since $|\mathrm Q|=1$, the differential acts on a cochain $A$, viewed as an operator on the Chevalley-Eilenberg complex, as
    \begin{equation}
        \{\mathrm Q,A\}=\mathrm QA-(-1)^{|A|}A\mathrm Q\,.
    \end{equation}
    For a cochain $A$, viewed as a multiplication operator in the odd variables, the bracket $\{\mathrm{Q},A\}$ contains no derivatives with respect to these variables and is therefore again an $\mathrm{End}(\mathcal{H})$-valued cochain. Indeed, all terms containing derivatives acting to the right cancel by the graded Leibniz rule.
\end{df}
An infinitesimal deformation of the $\mathfrak{Conf}_{2}$ module structure on $\mathcal{H}$ is encoded in a deformation of the differential $\mathrm{Q}$:
\begin{equation}
\begin{aligned}
\mathrm{Q}\rightarrow \mathrm{Q}+g\,\delta^{(1)}\mathrm{Q}\,,
\end{aligned}
\end{equation}
where $g$ is the deformation parameter. Here, $\delta^{(1)}\mathrm{Q}$ takes the following form:
\begin{equation}
    \label{cocycle_def}
    \delta^{(1)}\mathrm{Q}=\delta^{(1)}\mathcal{Q}+\delta^{(1)}\mathcal{\bar{Q}}=\sum_{m}c^{\,m}\delta^{(1)}L_{m}+\sum_{m}\bar{c}^{\,m}\delta^{(1)}\bar{L}_{m}\,,
\end{equation}
where $\delta^{(1)}L_{m}$ and $\delta^{(1)}\bar{L}_{m}$ are endomorphisms of the vector space $\mathcal{H}$. The nilpotency condition for the infinitesimally deformed differential reads:
\begin{equation}
\begin{aligned}
\label{cocycle_eq_0}
    \{\mathrm{Q},\delta^{(1)}\mathrm{Q}\}=\{\mathcal{Q},\delta^{(1)}\mathcal{Q}\}+\{\bar{\mathcal{Q}},\delta^{(1)}\bar{\mathcal{Q}}\}\,+\\+\,\{\bar{\mathcal{Q}},\delta^{(1)}\mathcal{Q}\}+\{\mathcal{Q},\delta^{(1)}\bar{\mathcal{Q}}\}=0\,.
\end{aligned}
\end{equation}
This condition means that $\delta^{(1)}\mathrm{Q}$ is a $\{\mathrm{Q},\cdot\}$-\emph{cocycle}. It may occur, however, that the first-order deformed representation is \emph{intertwined} with the original one, rendering the deformation trivial. This corresponds to the situation where the cocycle $\delta^{(1)}\mathrm{Q}$ is $\{\mathrm{Q},\cdot\}$-\emph{exact}:
\begin{equation}
    \delta^{(1)}\mathrm{Q}=\{\mathrm{Q},S\}=\{\mathcal{Q},S\}+\{\bar{\mathcal{Q}},S\}\,.
\end{equation}
Here, $S$ is a $0$-cochain in the adjoint Chevalley-Eilenberg complex, or simply an element of $\mathrm{End}(\mathcal{H})$. Thus, the infinitesimal deformations of the conformal module structure on the space $\mathcal{H}$ are given by the first cohomology of the differential $\{\mathrm{Q},\cdot\}$.

It is useful to rewrite these equations in components. The condition that $\delta^{(1)}\mathrm{Q}$ is a cocycle is equivalent to an infinite system of equations for the corrections $\delta^{(1)}L_{m}$ and $\delta^{(1)}\bar{L}_{m}$:
\begin{equation}
\begin{aligned}
    \label{cocycle_eq}
    [L_{m},\delta^{(1)}L_{n}]+[\delta^{(1)}L_{m},L_{n}]-(m-n)\,\delta^{(1)}L_{m+n}=0\,,\\
    [\bar{L}_{m},\delta^{(1)}\bar{L}_{n}]+[\delta^{(1)}\bar{L}_{m},\bar{L}_{n}]-(m-n)\,\delta^{(1)}\bar{L}_{m+n}=0\,,\\
    [L_{m},\delta^{(1)}\bar{L}_{n}]+[\delta^{(1)}L_{m},\bar{L}_{n}]=0\,.
\end{aligned}
\end{equation}
The exactness condition for the cocycle reads as follows:
\begin{eqnarray}
\label{exact_cocycle}
    \delta^{(1)}L_{m}=[L_{m},S]\,, \quad\delta^{(1)}\bar{L}_{m}=[\bar{L}_{m},S]\,.
\end{eqnarray}

Now that the basic notation has been established, we can formulate our first key statement: marginal observables of a two-dimensional CFT correspond to certain classes in the \emph{first cohomology} of the operator $\{\mathrm{Q},\cdot\}$. The converse, generally speaking, does not hold, as there exist examples of non-marginal deformations of CFT, such as the dilaton deformation \cite{Chodos:1973gt}. In this paper, we discuss a particular case of marginal deformations, namely, so-called \emph{current-current} deformations of a CFT \cite{Dashen:1974hp, Gerganov:2000mt}. The general cocycles associated with these deformations are considered in detail in \textbf{Section~\ref{subsec: Marginal_cocycles}}. The most illustrative example is discussed in \textbf{Section~\ref{subsec: free_boson_example}}.

In quantum field theory, marginal deformations are generally obstructed by the \emph{conformal anomaly}. While the theory remains conformal at first order in perturbation theory, at second order, quantum effects give rise to a non-vanishing quadratic term of the \emph{beta function} \cite{Cardy:1996xt}. At the level of the conformal algebra representation, this comes from the cohomological obstruction to deforming the $\mathfrak{Conf}_{2}$ module structure on $\mathcal{H}$.

Namely, consider the extension of the marginal deformation to higher orders of perturbation theory:
\begin{equation}
     \mathrm{Q}\rightarrow \mathrm{Q}+g\,\delta^{(1)}\mathrm{Q}+g^{2}\,\delta^{(2)}\mathrm{Q}+g^{3}\,\delta^{(3)}\mathrm{Q}+...
\end{equation}
In particular, $\delta^{(2)}\mathrm{Q}$ satisfies the \emph{Maurer-Cartan} equation:
\begin{equation}
\label{Maurer-Cartan_eq}
    \{\mathrm{Q},\delta^{(2)}\mathrm{Q}\}+\frac{1}{2}\{\delta^{(1)}\mathrm{Q},\delta^{(1)}\mathrm{Q}\}=0\,.
\end{equation}
The obstruction to the existence of $\delta^{(2)}\mathrm{Q}$ is given by the cohomology class of the $2$-cocycle $\frac{1}{2}\{\delta^{(1)}\mathrm{Q},\delta^{(1)}\mathrm{Q}\}$ in $H_{\{\mathrm{Q},\cdot \}}^{2}(\mathfrak{Conf}_{2},\mathrm{End}(\mathcal{H}))$. In \textbf{Section~\ref{subsec: Cardy_formula}}, we explicitly show that for current-current deformations, this cohomology class is proportional to the \emph{Cardy formula} \cite{Cardy:1996xt}, which determines the quadratic contribution to the beta function in conformal perturbation theory.

\section{Beta function as cohomological obstruction}
\subsection{General $J\bar{J}$-cocycles}
\label{subsec: Marginal_cocycles}
A special case of marginal CFT deformations is provided by the so-called \emph{current-current} deformations \cite{Dashen:1974hp, Gerganov:2000mt}.
\begin{df}
    Given a Lie algebra $\mathfrak{g}$ equipped with an invariant symmetric bilinear form $\eta$, the associated current algebra $\mathfrak{Cur}_{\mathfrak{g},\eta}$ is defined as follows. Consider a basis $e_{\alpha}$ of $\mathfrak{g}$. Let $f_{\alpha\beta}^{\gamma}$ denote the structure constants of $\mathfrak{g}$ in this basis, and let $\eta_{\alpha\beta}$ be the matrix elements of the form $\eta$. Then the generators $J_{\alpha(m)}$ of the algebra $\mathfrak{Cur}_{\mathfrak{g},\eta}$ satisfy the following commutation relations:
    \begin{equation}
    \label{Cur_algebra}
    [J_{\alpha(m)},J_{\beta(n)}]=f_{\alpha\beta}^{\gamma} \,J_{\gamma(m+n)}+\eta_{\alpha\beta}\cdot m\cdot \delta_{m+n,0}\,.
    \end{equation}
\end{df}
\begin{rem}
    The underlying Lie algebra $\mathfrak{g}$ is embedded into $\mathfrak{Cur}_{\mathfrak{g},\eta}$ as the zero-mode subalgebra spanned by $J_{\alpha(0)}$. If this zero-mode algebra is simple, the invariant form $\eta$ can be chosen proportional to the Killing form. In this case, $\mathfrak{Cur}_{\mathfrak{g},\eta}$ coincides with the standard affine Kac-Moody algebra \cite{Kac:1987gg}.
\end{rem}
Suppose that the state space $\mathcal{H}$, in addition to being a representation space for the conformal algebra $\mathfrak{Conf}_{2}$, also carries a representation of two copies of the current algebra $\mathfrak{Cur}_{\mathfrak{g},\eta}\oplus \overline{\mathfrak{Cur}}_{\bar{\mathfrak{g}},\bar{\eta}}$. In general, the Lie algebra $\mathfrak{g}$ need not coincide with its anti-chiral counterpart $\bar{\mathfrak{g}}$. The relations between the current and conformal algebra generators take the following form:
\begin{equation}
\label{primary_fields}
\begin{aligned}
    [L_{m},J_{\alpha(n)}]=-n\,J_{\alpha(m+n)}\,,\quad [L_{m},\bar{J}_{\bar{\alpha}(n)}]=0\,,\\
    [\bar{L}_{m},\bar{J}_{\bar{\alpha}(n)}]=-n\,\bar{J}_{\bar{\alpha}(m+n)}\,,\quad [\bar{L}_{m},J_{\alpha(n)}]=0\,.
\end{aligned}
\end{equation}

We now list a set of fairly general assumptions on the CFT state space $\mathcal H$ that we will impose throughout the remainder of this paper.

\begin{itemize}
\setlength{\itemsep}{10pt}
    \item \textbf{Grading}. The state space $\mathcal{H}$ admits a direct sum decomposition
    \begin{equation}
    \label{decomposition}
    \begin{aligned}
    \mathcal H=\bigoplus_{h,\bar h\geq0}\mathcal H_{(h,\bar h)}\,,
    \end{aligned}
    \end{equation}
    such that $H_{(h,\bar h)}$ are the eigenspaces for $L_{0}$ and $\bar{L}_{0}$:
    \begin{equation}
    \begin{aligned}
        L_0|_{\mathcal H_{(h,\bar h)}}=h\,\mathrm{id}_{\mathcal H_{(h,\bar h)}}\,, \qquad \bar L_0|_{\mathcal H_{(h,\bar h)}}=\bar h\,\mathrm{id}_{\mathcal H_{(h,\bar h)}}\,.
    \end{aligned}
    \end{equation}
    Every vector has finitely many nonzero weight components.
    \item \textbf{Unitarity}. The space $\mathcal{H}$ carries a positive-definite Hermitian inner product $\braket{\cdot,\cdot}:\overline{\mathcal{H}}\otimes\mathcal{H}\rightarrow\mathbb{C}$ satisfying
    \begin{equation}
    \begin{aligned}
    L_m^\dagger=L_{-m}\,,\qquad
    \bar L_m^\dagger=\bar L_{-m}\,.
    \end{aligned}
    \end{equation}
    In Hermitian bases of the current algebras,
    \begin{equation}
    \begin{aligned}
    J_{\alpha(m)}^{\,\dagger}=J_{\alpha(-m)}\,,\qquad
    \bar J_{\bar\alpha(m)}^{\,\dagger}=\bar J_{\bar\alpha(-m)}\,.
    \end{aligned}
    \end{equation}
    \item \textbf{Vacuum}. There exists a vacuum vector $\ket{\omega}\in\mathcal H_{(0,0)}$ such that
    \begin{equation}
    \begin{aligned}
    L_{n\geq-1}\ket{\omega}=\bar L_{n\geq-1}\ket{\omega}=0\,,\\
    J_{\alpha(n\geq 0)}\ket{\omega}
    =\bar J_{\bar\alpha(n\geq 0)}\ket{\omega}=0\,.
    \end{aligned}
    \end{equation}
    \item \textbf{Truncation}. For any state $\ket{\phi}\in\mathcal{H}$, we assume that
    \begin{equation}
    L_{m}\ket{\phi}=\bar{L}_{m}\ket{\phi}=J_{\alpha(m)}\ket{\phi} = \bar{J}_{\bar{\alpha}(m)}\ket{\phi} = 0
    \end{equation}
    for sufficiently large positive $m$.
\end{itemize}
For example, these assumptions hold in unitary WZW models on compact semisimple groups \cite{Eberhardt2019WZW}.
\begin{rem}
    It is crucial to emphasize that we do not assume the conformal algebra generators to be constructed from $J_{\alpha(m)}$ and $\bar{J}_{\bar{\alpha}(m)}$, as is the case in the Sugawara construction \cite{Sugawara:1967rw}. Instead, we merely assume that the CFT under consideration contains the chiral and anti-chiral current algebras.
\end{rem}

Early analyses of the deformation of the CFT stress-energy tensor under marginal perturbations were carried out in the context of conformal and string field theory \cite{campbell_stress_1991,sen_background_1990}. Motivated by their arguments, we arrive at the definition of cocycles governing the deformation of the conformal algebra module structure on the space of states $\mathcal{H}$ corresponding to a current-current deformation of the CFT, as formulated below.
\begin{df}
    We define the $J\bar{J}$-cocycles as
    \begin{equation}
        \label{curcur_cocycle}
        \delta^{(1)}\mathrm{Q}_{\alpha\bar{\alpha}}=\sum_{m}c^{\,m}J_{\alpha(m)}\bar{J}_{\bar{\alpha}(0)}+\sum_{m}\bar{c}^{\,m}J_{\alpha(0)}\bar{J}_{\bar{\alpha}(m)}\,.
    \end{equation}
    Using relations \eqref{primary_fields}, it is straightforward to show that it is indeed a $\{\mathrm{Q},\cdot\}$-cocycle. Thus, this cocycle defines an infinitesimal deformation of $\mathrm{Q}$:
    \begin{equation}
        \mathrm{Q}\rightarrow \mathrm{Q}+g^{\alpha\bar{\alpha}}\delta^{(1)}\mathrm{Q}_{\alpha\bar{\alpha}}\,.
    \end{equation}
    Here, $g^{\alpha\bar{\alpha}}$ denote the deformation parameters. In terms of the deformations of the conformal algebra generators, this reads:
    \begin{equation}
    \label{cocycle_eq_1}
    L_{m}\rightarrow L_{m}+g^{\alpha\bar{\alpha}}J_{\alpha(m)}\bar{J}_{\bar{\alpha}(0)}\,,\quad \bar{L}_{m}\rightarrow \bar{L}_{m}+g^{\alpha\bar{\alpha}}J_{\alpha(0)}\bar{J}_{\bar{\alpha}(m)}\,.
    \end{equation}
\end{df}
\begin{rem}
    Strictly speaking, the cocycle considered in \cite{campbell_stress_1991,sen_background_1990} has the form
    \begin{equation}
        \label{Sen-Nelson}
        \delta^{(1)}\mathrm{Q}_{\alpha\bar{\alpha}}=\sum_{m}c^{\,m}\left(\sum_{k\in\mathbb{Z}}J_{\alpha(m+k)}\bar{J}_{\bar{\alpha}(k)}\right)+\sum_{m}\bar{c}^{\,m}\left(\sum_{k\in\mathbb{Z}}J_{\alpha(k)}\bar{J}_{\bar{\alpha}(m+k)}\right)\,.
    \end{equation}
    This expression should be understood as an element of the CE complex with values in $\mathrm{End}(\mathcal{H})^{*}$, where the star denotes the completion with respect to the weak topology. However, one can show that within the complex $\mathrm{CE}^{\bullet}(\mathfrak{Conf}_{2},\mathrm{End}(\mathcal{H})^{*})$, the cocycles \eqref{curcur_cocycle} and \eqref{Sen-Nelson} are equivalent.
\end{rem}
\begin{st}
Suppose that $J_{\alpha(0)}\bar{J}_{\bar{\alpha}(0)}\neq 0$. Then, the cocycle $\delta^{(1)}\mathrm{Q}_{\alpha\bar{\alpha}}$ is not an exact cocycle.
\end{st}
\begin{pr}
    We prove this by contradiction. According to \eqref{exact_cocycle}, exactness of the cocycle $\delta^{(1)}\mathrm{Q}_{\alpha\bar{\alpha}}$ would imply the existence of an operator $S_{\alpha\bar{\alpha}}$ such that
    \begin{equation}
        J_{\alpha(m)}\bar{J}_{\bar{\alpha}(0)}=[L_{m},S_{\alpha\bar{\alpha}}]\,.
    \end{equation}
    Suppose there exists an operator $S_{\alpha\bar{\alpha}}$ such that $J_{\alpha(0)}\bar{J}_{\bar{\alpha}(0)} = [L_0, S_{\alpha\bar{\alpha}}]$. Consider $\ket{\phi} \in \mathcal{H}$ such that $L_0 \ket{\phi} = h_\phi \ket{\phi}$. Then
\begin{eqnarray}
\notag\bra{\phi} J_{\alpha(0)}^{\,\dagger} \bar{J}^{\,\dagger}_{\bar{\alpha}(0)}\bar{J}_{\bar{\alpha}(0)}J_{\alpha(0)} \ket{\phi} = \bra{\phi} J_{\alpha(0)}^{\,\dagger}\bar{J}^{\,\dagger}_{\bar{\alpha}(0)} [L_0, S_{\alpha\bar{\alpha}}] \ket{\phi} =\\= \bra{\phi} J_{\alpha(0)}^{\,\dagger} \bar{J}^{\,\dagger}_{\bar{\alpha}(0)}L_0 S_{\alpha\bar{\alpha}} \ket{\phi} - h_\phi \cdot \bra{\phi} J_{\alpha(0)}^{\,\dagger} \bar{J}^{\,\dagger}_{\bar{\alpha}(0)}S_{\alpha\bar{\alpha}} \ket{\phi}.
\end{eqnarray}
Using that $L_0 = L_0^{\dagger}$, we obtain
\begin{eqnarray}
\notag\bra{\phi} J_{\alpha(0)}^{\,\dagger} \bar{J}^{\,\dagger}_{\bar{\alpha}(0)}L_0 S_{\alpha\bar{\alpha}} \ket{\phi} - h_\phi \cdot \bra{\phi} J_{\alpha(0)}^{\dagger}\bar{J}^{\dagger}_{\bar{\alpha}(0)}S_{\alpha\bar{\alpha}} \ket{\phi} =\\= h_\phi \cdot (\bra{\phi} J_{\alpha(0)}^{\,\dagger}\bar{J}^{\,\dagger}_{\bar{\alpha}(0)} S_{\alpha\bar{\alpha}} \ket{\phi} - \bra{\phi} J_{\alpha(0)}^{\,\dagger}\bar{J}^{\,\dagger}_{\bar{\alpha}(0)} S_{\alpha\bar{\alpha}} \ket{\phi})=0\,.
\end{eqnarray}
Since the inner product was assumed to be positive-definite, it follows that $J_{\alpha(0)}\bar{J}_{\bar{\alpha}(0)}$ acts as zero on the state space $\mathcal{H}$. This is a contradiction.
\end{pr}
\subsection{Example: free boson}
\label{subsec: free_boson_example}
Let us consider a key example: a free boson whose target space is a circle of radius $R$. In this case, the algebras of chiral and anti-chiral currents are abelian:
\begin{eqnarray}
    [J_{(m)},J_{(n)}]=[\bar{J}_{(m)},\bar{J}_{(n)}]=m\cdot\delta_{m,-n}\,.
\end{eqnarray}
The corresponding representation of the conformal algebra is given by
\begin{equation}
\mathcal{H}=\bigoplus_{m,n\in\mathbb{Z}}\mathcal{V}_{m,n}\otimes \overline{\mathcal{V}}_{m,n}\,.
\end{equation}
The space $\mathcal{V}_{m,n} \otimes \overline{\mathcal{V}}_{m,n}$ is generated by the action of the creation operators $J_{(-k)}$ and $\bar{J}_{(-k)}$ ($k > 0$) on the highest-weight vector $|m,n\rangle$. The zero modes, in turn, act as follows:
\begin{equation}
\label{zero_modes}
    J_{(0)}\ket{m,n}=\left(\frac{m}{R}+\frac{nR}{2}\right)\ket{m,n}\,,\quad \bar{J}_{(0)}\ket{m,n}=\left(\frac{m}{R}-\frac{nR}{2}\right)\ket{m,n}\,.
\end{equation}
The conformal generators are given by the following expressions \cite{diFrancesco1997conformal,Sugawara:1967rw}:
\begin{equation}
\label{Sugawara}
\begin{aligned}
     L_{0}=\frac{1}{2}J_{(0)}^{\,2}+\sum_{n>0}J_{(-n)}J_{(n)}\,,\quad  L_{m\neq0}=\frac{1}{2}\sum_{n\in\mathbb{Z}}J_{(-n)}J_{(n+m)}\,,\\
     \bar{L}_{0}=\frac{1}{2}\bar{J}_{(0)}^{\,2}+\sum_{n>0}\bar{J}_{(-n)}\bar{J}_{(n)}\,,\quad  \bar{L}_{m\neq0}=\frac{1}{2}\sum_{n\in\mathbb{Z}}\bar{J}_{(-n)}\bar{J}_{(n+m)}\,.
     \end{aligned}
\end{equation}
Consider a deformation induced by an infinitesimal dilatation of the radius:
\begin{equation}
    R\rightarrow (1-g)\cdot R\,.
\end{equation}
Note that, since $R$ appears only in the zero-mode expressions \eqref{zero_modes}, this rescaling is realized by the following transformation:
\begin{equation}
    \notag J_{(0)}\rightarrow J_{(0)}+g\cdot\bar{J}_{(0)}\,,\quad \bar{J}_{(0)}\rightarrow\bar{J}_{(0)}+g\cdot J_{(0)}\,.
\end{equation}
It is worth emphasizing a striking feature of this formula: the correction to the chiral current modes explicitly involves the modes of the anti-chiral currents. Such a deformation of the zero modes induces a deformation of the conformal algebra generators, coinciding exactly with \eqref{cocycle_eq_1}:
\begin{equation}
\label{min_cocycle_example}
    L_{m}\rightarrow L_{m}+g\cdot J_{(m)}\bar{J}_{(0)}\,,\quad \bar{L}_{m}\rightarrow \bar{L}_{m}+g\cdot J_{(0)}\bar{J}_{(m)}\,.
    \end{equation}
Clearly, this deformation is non-trivial. Furthermore, in the case at hand, it is free of cohomological obstructions at higher orders and can be extended to all orders in perturbation theory. Indeed, the radius scaling induces the following transformation of $J_{(0)}$ and $\bar{J}_{(0)}$ to all orders:
\begin{equation}
\begin{aligned}
    J_{(0)}\rightarrow J_{(0)}+g\cdot \bar{J}_{(0)}+\sum_{k\geq 2}\left(\frac{J_{(0)}+\bar{J}_{(0)}}{2}\right)\cdot g^{k}\,,\\
    \bar{J}_{(0)}\rightarrow \bar{J}_{(0)}+g\cdot J_{(0)}+\sum_{k\geq 2}\left(\frac{\bar{J}_{(0)}+J_{(0)}}{2}\right)\cdot g^{k}\,.
\end{aligned}
\end{equation}
Substituting the deformed $J_{(0)}$ and $\bar{J}_{(0)}$ into the Sugawara construction \eqref{Sugawara}, we obtain corrections to the conformal generators to all orders in $g$.
\begin{rem}
    It is natural to ask what such a deformation corresponds to in the case of a free boson with a non-compact target space. Due to the presence of a position zero mode $q$ satisfying
\begin{equation}
    [q, J_{(m)}] = [q, \bar{J}_{(m)}] = \delta_{m,0}\,,
\end{equation}
the cocycle \eqref{min_cocycle_example} becomes \emph{exact}, and the corresponding deformation is trivial. This does not contradict the general arguments presented in the proof of \textbf{Proposition 1}, as the free boson theory with target space $\mathbb{R}$ involves \emph{non-normalizable} highest-weight vectors \cite{diFrancesco1997conformal}.
\end{rem}
\subsection{First obstruction and Cardy formula}
\label{subsec: Cardy_formula}
In this section, we compute the general form of the cocycle whose class constitutes a cohomological obstruction to the $J\bar{J}$-deformation. We will show that this class encodes the second order coefficient of the perturbative \emph{beta-function} \cite{Cardy:1996xt}.

Thus, we consider a deformation of the conformal algebra module $\mathcal{H}$ along the direction
\begin{equation}
\delta^{(1)}\mathrm{Q}=g^{\alpha\bar{\alpha}}\delta^{(1)}\mathrm{Q}_{\alpha\bar{\alpha}}\,,
\end{equation}
where $\delta^{(1)}\mathrm{Q}_{\alpha\bar{\alpha}}$ is given by \eqref{curcur_cocycle} and $g^{\alpha\bar{\alpha}}$ denote the deformation parameters. In contrast to the notation introduced in \textbf{Section 1.2}, here the deformation parameters are absorbed into the definition of $\delta^{(1)}\mathrm{Q}$. Then, according to equation \eqref{Maurer-Cartan_eq}, the cohomological obstruction to the deformation along $\delta^{(1)}\mathrm{Q}$ is given by the cohomology class of $\frac{1}{2}\{\delta^{(1)}\mathrm{Q},\delta^{(1)}\mathrm{Q}\}$.

The cocycle representing the potential obstruction is given by
\begin{equation}
\begin{aligned}
    \frac{1}{2}\{\delta^{(1)}\mathrm{Q},\delta^{(1)}\mathrm{Q}\}=\sum_{m<n}c^{\,m}c^{\,n}\cdot[\delta^{(1)}L_{m},\delta^{(1)}L_{n}]\,+\\+\,\sum_{m<n}\bar{c}^{\,m}\bar{c}^{\,n}\cdot[\delta^{(1)}\bar{L}_{m},\delta^{(1)}\bar{L}_{n}]+\sum_{m,n}c^{\,m}\bar{c}^{\,n}[\delta^{(1)}L_{m},\delta^{(1)}\bar{L}_{n}]\,.
    \end{aligned}
\end{equation}
Then equation \eqref{Maurer-Cartan_eq} for $\delta^{(2)}\mathrm{Q}$ in components takes the form:
\begin{equation}
\begin{aligned}
\label{Maurer_Cartan_components_1}
    [L_{m},\delta^{(2)}L_{n}]+[\delta^{(2)}L_{m},L_{n}]-(m-n)\,\delta^{(2)}L_{m+n}-\\\ -\,\delta_{m+n,0}\cdot \frac{m^{3}-m}{12}\cdot\delta^{(2)}C+[\delta^{(1)}L_{m},\delta^{(1)}L_{n}]=0\,,
\end{aligned}
\end{equation}
\begin{equation}
\begin{aligned}
\label{Maurer_Cartan_components_2}
    [\bar{L}_{m},\delta^{(2)}\bar{L}_{n}]+[\delta^{(2)}\bar{L}_{m},\bar{L}_{n}]-(m-n)\,\delta^{(2)}\bar{L}_{m+n}-\\\ -\,\delta_{m+n,0}\cdot \frac{m^{3}-m}{12}\cdot\delta^{(2)}\bar{C}+[\delta^{(1)}\bar{L}_{m},\delta^{(1)}\bar{L}_{n}]=0\,,
\end{aligned}
\end{equation}
\begin{equation}
\begin{aligned}
\label{Maurer_Cartan_components_3}
    [L_{m},\delta^{(2)}\bar{L}_{n}]+[\delta^{(2)}L_{m},\bar{L}_{n}]+[\delta^{(1)}L_{m},\delta^{(1)}\bar{L}_{n}]=0\,.
\end{aligned}
\end{equation}
Our goal now is to evaluate the commutators $[\delta^{(1)}L_{m},\delta^{(1)}L_{n}],\,[\delta^{(1)}\bar{L}_{m},\delta^{(1)}\bar{L}_{n}]$ and $[\delta^{(1)}L_{m},\delta^{(1)}\bar{L}_{n}]$ for $\delta^{(1)}L_{m}=g^{\alpha\bar{\alpha}}J_{\alpha(m)}\bar{J}_{\bar{\alpha}(0)}$ and $\delta^{(1)}\bar{L}_{m}=g^{\alpha\bar{\alpha}}J_{\alpha(0)}\bar{J}_{\bar{\alpha}(m)}$.
\begin{st}
The commutators are given by the following formulas:
\begin{equation}
\begin{aligned}
[\delta^{(1)}L_{m},\delta^{(1)}L_{n}]=\frac{1}{2}g^{\alpha\bar{\alpha}}g^{\beta\bar{\beta}}\bar{f}_{\bar{\alpha}\bar{\beta}}^{\bar{\gamma}}\bar{J}_{\bar{\gamma}(0)}(J_{\beta(n)}J_{\alpha(m)}-J_{\beta(m)}J_{\alpha(n)})\,+\\+\,(m-n)\cdot \frac{1}{2}g^{\alpha\bar{\alpha}}g^{\beta\bar{\beta}}\eta_{\alpha\beta}\bar{J}_{\bar{\alpha}(0)}\bar{J}_{\bar{\beta}(0)}\cdot \delta_{m+n,0}\,,
\end{aligned}
\end{equation}
\begin{equation}
\begin{aligned}
[\delta^{(1)}\bar{L}_{m},\delta^{(1)}\bar{L}_{n}]=\frac{1}{2}g^{\alpha\bar{\alpha}}g^{\beta\bar{\beta}}f_{\alpha\beta}^{\gamma}J_{\gamma(0)}(\bar{J}_{\bar{\beta}(n)}\bar{J}_{\bar{\alpha}(m)}-\bar{J}_{\bar{\beta}(m)}\bar{J}_{\bar{\alpha}(n)})\,+\\+\,(m-n)\cdot \frac{1}{2}g^{\alpha\bar{\alpha}}g^{\beta\bar{\beta}}\bar{\eta}_{\bar{\alpha}\bar{\beta}}J_{\alpha(0)}J_{\beta(0)}\cdot \delta_{m+n,0}\,,
\end{aligned}
\end{equation}
\begin{equation}
\begin{aligned}
[\delta^{(1)}L_{m},\delta^{(1)}\bar{L}_{n}]=\frac{1}{2}g^{\alpha\bar{\alpha}}g^{\beta\bar{\beta}}\bar{f}_{\bar{\alpha}\bar{\beta}}^{\bar{\gamma}}\bar{J}_{\bar{\gamma}(n)}(J_{\beta(0)}J_{\alpha(m)}-J_{\beta(m)}J_{\alpha(0)})\,+\\+\,\frac{1}{2}g^{\alpha\bar{\alpha}}g^{\beta\bar{\beta}}f_{\alpha\beta}^{\gamma}J_{\gamma(m)}(\bar{J}_{\bar{\alpha}(0)}\bar{J}_{\bar{\beta}(n)}-\bar{J}_{\bar{\alpha}(n)}\bar{J}_{\bar{\beta}(0)})\,.
\end{aligned}
\end{equation}
\end{st}
\begin{pr}
This follows from a direct calculation employing the current algebra commutation relations \eqref{Cur_algebra} and their antichiral counterparts.
\end{pr}
\begin{st}
The terms containing $\eta_{\alpha\beta}$ and $\bar{\eta}_{\bar{\alpha}\bar{\beta}}$ form an exact 2-cocycle
\begin{equation}
\begin{aligned}
    \alpha_{2} = \frac{1}{2}g^{\alpha\bar{\alpha}}g^{\beta\bar{\beta}}\eta_{\alpha\beta}\sum_{m<n}c^{\,m}c^{\,n}(m-n)\bar{J}_{\bar{\alpha}(0)}\bar{J}_{\bar{\beta}(0)}\cdot \delta_{m+n,0}\,+\\+\,\frac{1}{2}g^{\alpha\bar{\alpha}}g^{\beta\bar{\beta}}\bar{\eta}_{\bar{\alpha}\bar{\beta}}\sum_{m<n}\bar{c}^{\,m}\bar{c}^{\,n}(m-n)J_{\alpha(0)}J_{\beta(0)}\cdot \delta_{m+n,0}\,.
\end{aligned}
\end{equation}
This implies that they can be eliminated from equations \eqref{Maurer_Cartan_components_1}, \eqref{Maurer_Cartan_components_2}, and \eqref{Maurer_Cartan_components_3} by a shift of the second order corrections $\delta^{(2)}L_{m}$ and $\delta^{(2)}\bar{L}_{m}$.
\end{st}
\begin{pr}
Indeed, these terms are eliminated from the equations by the shift
\begin{equation}
\begin{aligned}
\delta^{(2)}L_{m}\rightarrow \delta^{(2)}L_{m}+ \frac{1}{2}g^{\alpha\bar{\alpha}}g^{\beta\bar{\beta}}\eta_{\alpha\beta}\bar{J}_{\bar{\alpha}(0)}\bar{J}_{\bar{\beta}(0)}\cdot \delta_{m,0}\,,\\\delta^{(2)}\bar{L}_{m}\rightarrow \delta^{(2)}\bar{L}_{m}+ \frac{1}{2}g^{\alpha\bar{\alpha}}g^{\beta\bar{\beta}}\bar{\eta}_{\bar{\alpha}\bar{\beta}}J_{\alpha(0)}J_{\beta(0)}\cdot \delta_{m,0}\,.
\end{aligned}
\end{equation}
\end{pr}
\begin{st}
The $2$-cocycle $\frac{1}{2}\{\delta^{(1)}\mathrm{Q},\delta^{(1)}\mathrm{Q}\}$, where $\delta^{(1)}\mathrm{Q}$ is the $J\bar{J}$-cocycle \eqref{curcur_cocycle}, is cohomologically equivalent to the cocycle
\begin{equation}
\begin{aligned}
\beta_{2}=\frac{1}{2}g^{\alpha\bar{\alpha}}g^{\beta\bar{\beta}} f_{\alpha\beta}^{\gamma}\bar{f}_{\bar{\alpha}\bar{\beta}}^{\,\bar{\gamma}}\cdot \sum_{m<n}c^{\,m} c^{\,n}\cdot (m-n)\cdot J_{\gamma(m+n)}\bar{J}_{\bar{\gamma}(0)}\,+\\+\,\frac{1}{2}g^{\alpha\bar{\alpha}}g^{\beta\bar{\beta}} f_{\alpha\beta}^{\gamma}\bar{f}_{\bar{\alpha}\bar{\beta}}^{\,\bar{\gamma}}\cdot \sum_{m<n}\bar{c}^{\,m} \bar{c}^{\,n}\cdot (m-n)\cdot J_{\gamma(0)}\bar{J}_{\bar{\gamma}(m+n)}\,+\\+\,\frac{1}{2}g^{\alpha\bar{\alpha}}g^{\beta\bar{\beta}} f_{\alpha\beta}^{\gamma}\bar{f}_{\bar{\alpha}\bar{\beta}}^{\,\bar{\gamma}}\cdot \sum_{m,n}c^{\,m} \bar{c}^{\,n}\cdot (m-n)\cdot J_{\gamma(m)}\bar{J}_{\bar{\gamma}(n)}\,.
\end{aligned}
\end{equation}
\end{st}
\begin{pr}
Let us denote the combination $\frac{1}{2}g^{\alpha\bar{\alpha}}g^{\beta\bar{\beta}} f_{\alpha\beta}^{\gamma}\bar{f}_{\bar{\alpha}\bar{\beta}}^{\,\bar{\gamma}}$ by $\beta_{2}^{\gamma\bar{\gamma}}$. To show that the cocycles are equivalent, one needs to find an exact 2-cocycle whose addition to $\frac{1}{2}\{\delta^{(1)}\mathrm{Q},\delta^{(1)}\mathrm{Q}\}$ yields $\beta_{2}$. Equivalently, one needs to shift $\delta^{(2)}L_{m}$ and $\delta^{(2)}\bar{L}_{m}$ so that equations \eqref{Maurer_Cartan_components_1}, \eqref{Maurer_Cartan_components_2}, and \eqref{Maurer_Cartan_components_3} take the form
\begin{equation}
\label{Maurer_Cartan_with_beta_1}
\begin{aligned}
    [L_{m},\delta^{(2)}L_{n}]+[\delta^{(2)}L_{m},L_{n}]-(m-n)\,\delta^{(2)}L_{m+n}-\\\ -\,\delta_{m+n,0}\cdot \frac{m^{3}-m}{12}\cdot\delta^{(2)}C+\beta_{2}^{\gamma\bar{\gamma}}\cdot(m-n)\cdot J_{\gamma(m+n)}\bar{J}_{\bar{\gamma}(0)}=0\,,
\end{aligned}
\end{equation}
\begin{equation}
\label{Maurer_Cartan_with_beta_2}
\begin{aligned}
    [\bar{L}_{m},\delta^{(2)}\bar{L}_{n}]+[\delta^{(2)}\bar{L}_{m},\bar{L}_{n}]-(m-n)\,\delta^{(2)}\bar{L}_{m+n}-\\\ -\,\delta_{m+n,0}\cdot \frac{m^{3}-m}{12}\cdot\delta^{(2)}\bar{C}+\beta_{2}^{\gamma\bar{\gamma}}\cdot(m-n)\cdot J_{\gamma(0)}\bar{J}_{\bar{\gamma}(m+n)}=0\,,
\end{aligned}
\end{equation}
\begin{equation}
\label{Maurer_Cartan_with_beta_3}
\begin{aligned}
    [L_{m},\delta^{(2)}\bar{L}_{n}]+[\delta^{(2)}L_{m},\bar{L}_{n}]+\beta_{2}^{\gamma\bar{\gamma}}\cdot(m-n)\cdot J_{\gamma(m)}\bar{J}_{\bar{\gamma}(n)}=0\,.
\end{aligned}
\end{equation}
First, according to \textbf{Proposition 3}, we shift $\delta^{(2)}L_{m}$ and $\delta^{(2)}\bar{L}_{m}$ to eliminate the terms containing $\eta_{\alpha\beta}$ and $\bar{\eta}_{\bar{\alpha}\bar{\beta}}$. Afterwards, we make the following shift:
\begin{equation}
\begin{aligned}
\label{chiral_shift}
    \delta^{(2)}L_{m}\rightarrow \delta^{(2)}L_{m}+g^{\alpha\bar{\alpha}}g^{\beta\bar{\beta}}f_{\alpha\beta}^{\gamma}\bar{f}_{\bar{\alpha}\bar{\beta}}^{\bar{\gamma}}\bar{J}_{\bar{\gamma}(0)}J_{\gamma(m)}\sum_{k=1}^{\Lambda}\frac{1}{k}\,+\\+\,\frac{1}{2}g^{\alpha\bar{\alpha}}g^{\beta\bar{\beta}}\bar{f}_{\bar{\alpha}\bar{\beta}}^{\bar{\gamma}}\bar{J}_{\bar{\gamma}(0)}\left(\sum_{k=-\Lambda}^{-1}+\sum_{k=1}^{\Lambda}\right) \frac{1}{k}J_{\beta(m+k)}J_{\alpha(-k)}\,-\\-\,\frac{1}{2}g^{\alpha\bar{\alpha}}g^{\beta\bar{\beta}}f_{\alpha\beta}^{\gamma}J_{\gamma(m)}\left(\sum_{k=-\Lambda}^{-1}+\sum_{k=1}^{\Lambda}\right)\frac{1}{k}\bar{J}_{\bar{\alpha}(k)}\bar{J}_{\bar{\beta}(-k)}\,,
\end{aligned}
\end{equation}
\begin{equation}
\begin{aligned}
\label{antichiral_shift}
    \delta^{(2)}\bar{L}_{n}\rightarrow \delta^{(2)}\bar{L}_{n}+g^{\alpha\bar{\alpha}}g^{\beta\bar{\beta}}f_{\alpha\beta}^{\gamma}\bar{f}_{\bar{\alpha}\bar{\beta}}^{\bar{\gamma}}\bar{J}_{\bar{\gamma}(n)}J_{\gamma(0)}\sum_{k=1}^{\Lambda}\frac{1}{k}\,+\\+\,\frac{1}{2}g^{\alpha\bar{\alpha}}g^{\beta\bar{\beta}}f_{\alpha\beta}^{\gamma}J_{\gamma(0)}\left(\sum_{k=-\Lambda}^{-1}+\sum_{k=1}^{\Lambda}\right)\frac{1}{k}\bar{J}_{\bar{\beta}(n+k)}\bar{J}_{\bar{\alpha}(-k)}\,-\\-\,\frac{1}{2}g^{\alpha\bar{\alpha}}g^{\beta\bar{\beta}}\bar{f}_{\bar{\alpha}\bar{\beta}}^{\bar{\gamma}}\bar{J}_{\bar{\gamma}(n)}\left(\sum_{k=-\Lambda}^{-1}+\sum_{k=1}^{\Lambda}\right)\frac{1}{k}J_{\alpha(k)}J_{\beta(-k)}\,.
\end{aligned}
\end{equation}
Here $\Lambda$ is an arbitrary positive integer. Thus, we add an exact 2-cocycle depending on $\Lambda$ to the cocycle $\frac{1}{2}\{\delta^{(1)}\mathrm{Q},\delta^{(1)}\mathrm{Q}\}$. Clearly, the cohomology class of this sum does not depend on $\Lambda$. After all these transformations, equations \eqref{Maurer_Cartan_components_1}, \eqref{Maurer_Cartan_components_2}, and \eqref{Maurer_Cartan_components_3} take the form
\begin{equation}
\begin{aligned}
    \notag[L_{m},\delta^{(2)}L_{n}]+[\delta^{(2)}L_{m},L_{n}]-(m-n)\,\delta^{(2)}L_{m+n}-\delta_{m+n,0}\cdot \frac{m^{3}-m}{12}\cdot\delta^{(2)}C\,+\\+\,\frac{1}{2}g^{\alpha\bar{\alpha}}g^{\beta\bar{\beta}}\bar{f}_{\bar{\alpha}\bar{\beta}}^{\bar{\gamma}}\bar{J}_{\bar{\gamma}(0)}\sum_{k=-\Lambda}^{\Lambda}(J_{\beta(n+k)}J_{\alpha(m-k)}-J_{\beta(m+k)}J_{\alpha(n-k)})=0\,,
\end{aligned}
\end{equation}
\begin{equation}
\begin{aligned}
\notag[\bar{L}_{m},\delta^{(2)}\bar{L}_{n}]+[\delta^{(2)}\bar{L}_{m},\bar{L}_{n}]-(m-n)\,\delta^{(2)}\bar{L}_{m+n}-\delta_{m+n,0}\cdot \frac{m^{3}-m}{12}\cdot\delta^{(2)}\bar{C}\,+\\+\,\frac{1}{2}g^{\alpha\bar{\alpha}}g^{\beta\bar{\beta}}f_{\alpha\beta}^{\gamma}J_{\gamma(0)}\sum_{k=-\Lambda}^{\Lambda}(\bar{J}_{\bar{\beta}(n+k)}\bar{J}_{\bar{\alpha}(m-k)}-\bar{J}_{\bar{\beta}(m+k)}\bar{J}_{\bar{\alpha}(n-k)})=0\,,
\end{aligned}
\end{equation}
\begin{equation}
\begin{aligned}
    \notag[L_{m},\delta^{(2)}\bar{L}_{n}]+[\delta^{(2)}L_{m},\bar{L}_{n}]\,+\\+\,\frac{1}{2}g^{\alpha\bar{\alpha}}g^{\beta\bar{\beta}}\bar{f}_{\bar{\alpha}\bar{\beta}}^{\bar{\gamma}}\bar{J}_{\bar{\gamma}(n)}\sum_{k=-\Lambda}^{\Lambda}(J_{\beta(0+k)}J_{\alpha(m-k)}-J_{\beta(m+k)}J_{\alpha(0-k)})\,+\\+\,\frac{1}{2}g^{\alpha\bar{\alpha}}g^{\beta\bar{\beta}}f_{\alpha\beta}^{\gamma}J_{\gamma(m)}\sum_{k=-\Lambda}^{\Lambda}(\bar{J}_{\bar{\alpha}(0+k)}\bar{J}_{\bar{\beta}(n-k)}-\bar{J}_{\bar{\alpha}(n+k)}\bar{J}_{\bar{\beta}(0-k)})=0\,.
\end{aligned}
\end{equation}
Now, using the fact that the cohomology class does not depend on $\Lambda$, let us consider the limit as $\Lambda \to \infty$. We compute this limit in the chiral equation; for the remaining ones, it is computed similarly. Specifically, let us consider the operator
\begin{equation}
\frac{1}{2}g^{\alpha\bar{\alpha}}g^{\beta\bar{\beta}}\bar{f}_{\bar{\alpha}\bar{\beta}}^{\bar{\gamma}}\bar{J}_{\bar{\gamma}(0)}\sum_{k=-\Lambda}^{\Lambda}(J_{\beta(n+k)}J_{\alpha(m-k)}-J_{\beta(m+k)}J_{\alpha(n-k)})
\end{equation}
and apply it to an arbitrary vector $\ket{\phi}$. We write this as follows:
\begin{equation}
\begin{aligned}
\frac{1}{2}g^{\alpha\bar{\alpha}}g^{\beta\bar{\beta}}\bar{f}_{\bar{\alpha}\bar{\beta}}^{\bar{\gamma}}\bar{J}_{\bar{\gamma}(0)}\sum_{k=-\Lambda}^{m}J_{\beta(n+k)}J_{\alpha(m-k)}\ket{\phi}\,+\\+\,\frac{1}{2}g^{\alpha\bar{\alpha}}g^{\beta\bar{\beta}}\bar{f}_{\bar{\alpha}\bar{\beta}}^{\bar{\gamma}}\bar{J}_{\bar{\gamma}(0)}\sum_{k=m+1}^{\Lambda}J_{\alpha(m-k)}J_{\beta(n+k)}\ket{\phi}\,-\\-\,\frac{1}{2}g^{\alpha\bar{\alpha}}g^{\beta\bar{\beta}}\bar{f}_{\bar{\alpha}\bar{\beta}}^{\bar{\gamma}}\bar{J}_{\bar{\gamma}(0)}\sum_{k=-\Lambda}^{n}J_{\beta(m+k)}J_{\alpha(n-k)}\ket{\phi}\,-\\-\,\frac{1}{2}g^{\alpha\bar{\alpha}}g^{\beta\bar{\beta}}\bar{f}_{\bar{\alpha}\bar{\beta}}^{\bar{\gamma}}\bar{J}_{\bar{\gamma}(0)}\sum_{k=n+1}^{\Lambda}J_{\alpha(n-k)}J_{\beta(m+k)}\ket{\phi}\,-\\-\,\frac{1}{2}g^{\alpha\bar{\alpha}}g^{\beta\bar{\beta}}f_{\alpha\beta}^{\gamma}\bar{f}_{\bar{\alpha}\bar{\beta}}^{\bar{\gamma}}\bar{J}_{\bar{\gamma}(0)}\sum_{k=m+1}^{\Lambda}J_{\gamma(m+n)}\ket{\phi}\,+\\+\,\frac{1}{2}g^{\alpha\bar{\alpha}}g^{\beta\bar{\beta}}f_{\alpha\beta}^{\gamma}\bar{f}_{\bar{\alpha}\bar{\beta}}^{\bar{\gamma}}\bar{J}_{\bar{\gamma}(0)}\sum_{k=n+1}^{\Lambda}J_{\gamma(m+n)}\ket{\phi}\,.
\end{aligned}
\end{equation}
Using the fact that $J_{\alpha(k)}\ket{\phi}=\bar{J}_{\bar{\alpha}(k)}\ket{\phi}=0$ for sufficiently large $k$, we find that the sums in the first four terms truncate in the limit $\Lambda \to \infty$ and cancel each other out. The remaining two terms combine into
\begin{equation}
\begin{aligned}
\left(\sum_{k=n+1}^{\Lambda}-\sum_{k=m+1}^{\Lambda}\right)\frac{1}{2}g^{\alpha\bar{\alpha}}g^{\beta\bar{\beta}}f_{\alpha\beta}^{\gamma}\bar{f}_{\bar{\alpha}\bar{\beta}}^{\bar{\gamma}}J_{\gamma(m+n)}\bar{J}_{\bar{\gamma}(0)}\ket{\phi}=\\=(\Lambda-n-\Lambda+m)\cdot \frac{1}{2}g^{\alpha\bar{\alpha}}g^{\beta\bar{\beta}}f_{\alpha\beta}^{\gamma}\bar{f}_{\bar{\alpha}\bar{\beta}}^{\bar{\gamma}}J_{\gamma(m+n)}\bar{J}_{\bar{\gamma}(0)}\ket{\phi}=\\=\frac{1}{2}g^{\alpha\bar{\alpha}}g^{\beta\bar{\beta}}f_{\alpha\beta}^{\gamma}\bar{f}_{\bar{\alpha}\bar{\beta}}^{\bar{\gamma}}\cdot (m-n)\cdot J_{\gamma(m+n)}\bar{J}_{\bar{\gamma}(0)}\ket{\phi}\,.
\end{aligned}
\end{equation} 
Computing analogous limits in the remaining two equations completes the proof of \textbf{Proposition 4}.
\end{pr}
\begin{rem}
    The cochains in \eqref{chiral_shift} and \eqref{antichiral_shift} also possess a limit as $\Lambda \to \infty$.
\end{rem}
We now turn to a discussion of our results. We have shown that the cocycle corresponding to the obstruction to the $J\bar{J}$-deformation is given by
\begin{equation}
\begin{aligned}
\label{obs_cocycle}
\beta_{2}=\frac{1}{2}g^{\alpha\bar{\alpha}}g^{\beta\bar{\beta}} f_{\alpha\beta}^{\gamma}\bar{f}_{\bar{\alpha}\bar{\beta}}^{\,\bar{\gamma}}\cdot \sum_{m<n}c^{\,m} c^{\,n}\cdot (m-n)\cdot J_{\gamma(m+n)}\bar{J}_{\bar{\gamma}(0)}\,+\\+\,\frac{1}{2}g^{\alpha\bar{\alpha}}g^{\beta\bar{\beta}} f_{\alpha\beta}^{\gamma}\bar{f}_{\bar{\alpha}\bar{\beta}}^{\,\bar{\gamma}}\cdot \sum_{m<n}\bar{c}^{\,m} \bar{c}^{\,n}\cdot (m-n)\cdot J_{\gamma(0)}\bar{J}_{\bar{\gamma}(m+n)}\,+\\+\,\frac{1}{2}g^{\alpha\bar{\alpha}}g^{\beta\bar{\beta}} f_{\alpha\beta}^{\gamma}\bar{f}_{\bar{\alpha}\bar{\beta}}^{\,\bar{\gamma}}\cdot \sum_{m,n}c^{\,m} \bar{c}^{\,n}\cdot (m-n)\cdot J_{\gamma(m)}\bar{J}_{\bar{\gamma}(n)}\,.
\end{aligned}
\end{equation}
Strikingly, this expression is proportional to a remarkable structure that coincides with the \emph{Cardy formula} for the leading contribution to the beta function:
\begin{equation}
    \beta_{2}^{\gamma\bar{\gamma}}(g)=\frac{1}{2}\,g^{\alpha\bar{\alpha}}g^{\beta\bar{\beta}}f_{\alpha\beta}^{\gamma}\bar{f}_{\bar{\alpha}\bar{\beta}}^{\bar{\gamma}}\,.
\end{equation}
From a field-theoretic point of view, it is well known that a non-vanishing beta function $\beta_{2}^{\gamma\bar{\gamma}}(g)$ constitutes an obstruction to preserving conformal symmetry of the deformed theory at second order in perturbation theory. Therefore, its appearance in the inhomogeneous terms of equations \eqref{Maurer_Cartan_with_beta_1}, \eqref{Maurer_Cartan_with_beta_2} and \eqref{Maurer_Cartan_with_beta_3} hints that this system does not have a solution. Indeed, the following propositions hold.
\begin{st}
    If the following $1$-cocycle is not $\{\mathcal{Q},\cdot\}$-exact
    \begin{equation}
    \label{Chodos-Thorn}
        \beta_{2}^{\gamma\bar{\gamma}}(g)\sum_{m}c^{\,m}\cdot m\cdot J_{\gamma(m)}\bar{J}_{\bar{\gamma}(0)}\,,
    \end{equation}
    then the $2$-cocycle $\beta_{2}$ is not $\{\mathrm{Q},\cdot\}$-exact.
\end{st}
\begin{pr}
    Consider the operator $b_{0}=\frac{\partial}{\partial c^{\,0}}$ of degree $-1$ on $\mathrm{CE}^{\bullet}(\mathfrak{Vir},\mathcal{H})$. Then
    \begin{equation}
        \{\mathcal{Q},b_{0}\}=L_{0}+\sum_{k}k\cdot c^{\,k}\frac{\partial}{\partial c^{\,k}}\,.
    \end{equation}
    Let us denote the right-hand side by $d$. Then, we notice that
    \begin{equation}
    \label{BRST_arg}
        \{\mathcal{Q},d\}=\{d,b_{0}\}=0\,.
    \end{equation}
    Using the fact that the state space $\mathcal{H}$ is a direct sum of \(L_0\)-eigenspaces, together with \eqref{BRST_arg}, we can explicitly decompose the complex \(\mathrm{CE}^{\bullet}(\mathfrak{Vir},\operatorname{End}(\mathcal H))\) into the kernel subcomplex \(\ker^\bullet(\{d,\cdot\})\) and an acyclic complement
    \begin{equation}
    \label{Vir_decomposition}
        \mathrm{CE}^{\bullet}(\mathfrak{Vir},\operatorname{End}(\mathcal H))= \ker^\bullet(\{d,\cdot\})\oplus \mathrm{Acyclic}^{\bullet}\,.
    \end{equation}
    Indeed, consider an \(r\)-cochain \(c^{m_1}...\,c^{m_r}S\), where \(S\in\mathrm{End}(\mathcal H)\). Then
    \begin{equation}
    \begin{aligned}
        \Pi_{\ker^r(\{d,\cdot\})}
        (c^{m_1}...\,c^{m_r}S)=c^{m_1}...\,c^{m_r} \sum_h \Pi_{h-m_1-...-m_r}\,S\,\Pi_h\,,
    \end{aligned}
    \end{equation}
    where \(\Pi_h\) denotes the projection onto the \(L_0\)-eigenspace with eigenvalue \(h\). The chiral component of $\beta_{2}$ lies in the component $\mathrm{ker}^{\bullet}(\{d,\cdot\})$. Now, we notice that within the subcomplex $\mathrm{ker}^{\bullet}(\{d,\cdot\})$, the following relation holds:
    \begin{equation}
        \{\mathcal{Q},\{b_{0},\cdot\}\}+\{b_{0},\{\mathcal{Q},\cdot\}\}=0\,.
    \end{equation}
    Thus, $\{b_{0},\cdot\}$ induces a well-defined map on the cohomology of $\{\mathcal{Q},\cdot\}$, as it maps exact cochains to exact ones and closed cochains to closed ones. 
    
    Now, suppose for a moment that the chiral component of the cocycle $\beta_{2}$ were $\{\mathcal{Q},\cdot\}$-exact. By the decomposition \eqref{Vir_decomposition}, we may choose its $\{\mathcal{Q},\cdot\}$-preimage to lie in \(\ker^\bullet(\{d,\cdot\})\). This would imply that the image of the chiral component of $\beta_{2}$ under $\{b_{0},\cdot\}$ must also be exact. On the other hand:
    \begin{equation}
    \label{ChodThor}
    \begin{aligned}
        \beta_{2}^{\gamma\bar{\gamma}}(g)\sum_{m,n}\{b_{0},c^{\,m} c^{\,n}\cdot (m-n)\}\cdot J_{\gamma(m+n)}\bar{J}_{\bar{\gamma}(0)}=\\=-2 \cdot \beta_{2}^{\gamma\bar{\gamma}}(g)\sum_{k}c^{k}\cdot k\cdot J_{\gamma(k)}\bar{J}_{\bar{\gamma}(0)}\,.
        \end{aligned}
    \end{equation}
    Thus, if the cocycle \eqref{ChodThor} is not \(\{\mathcal Q,\cdot\}\)-exact, neither is the chiral component of \(\beta_2\). Consequently, \(\beta_2\) is not an exact cocycle.
\end{pr}

\begin{rem}
    The cocycle \eqref{ChodThor} is precisely the Chodos-Thorn cocycle \cite{Chodos:1973gt} multiplied by the antichiral zero mode $\bar{J}_{\bar{\gamma}(0)}$.
\end{rem}

\begin{rem}
    The same argument could be applied to prove \textbf{Proposition 1}.
\end{rem}

\noindent It remains to show that the 1-cocycle \eqref{ChodThor} is not \(\{\mathcal Q,\cdot\}\)-exact. This leads us to our final proposition.

\begin{st}
    Assume that the invariant forms \(\eta\) and \(\bar\eta\) are positive definite and that the antichiral zero-mode algebra \(\bar{\mathfrak g}\) has trivial centre. Then, for any nonzero matrix \(\beta_2^{\gamma\bar\gamma}(g)\), the cocycle
    \begin{equation}
    \label{ChodThor2}
        \beta_2^{\gamma\bar\gamma}(g)
        \sum_k c^k\,k\,J_{\gamma(k)}\bar J_{\bar\gamma(0)}
    \end{equation}
    is not \(\{\mathcal Q,\cdot\}\)-exact.
\end{st}
\begin{pr}
    The proof is very similar to the proof of \textbf{Proposition 1}. Consider
    \begin{equation}
        \ket{\phi_{\bar{\beta}}}=\bar{J}_{\bar{\beta}(-1)}\ket{\omega}\,,\qquad\ket{\psi_{\bar{\beta}}}=\beta_2^{\alpha\bar\alpha}(g)\bar{f}_{\bar{\alpha}\bar{\beta}}^{\bar{\gamma}}\,J_{\alpha(-1)}\bar{J}_{\bar{\gamma}(-1)}\ket{\omega}\,.
    \end{equation}
    Since \(Z(\bar{\mathfrak g})=0\), there exists an index \(\bar\beta\) such that \(|\psi_{\bar\beta}\rangle\neq0\). For this value of \(\bar\beta\), the positive definiteness of the forms \(\eta\) and \(\bar\eta\) implies that
    \begin{equation}
        \braket{\psi_{\bar{\beta}}|\psi_{\bar{\beta}}}>0\,.
    \end{equation}
    It can be shown that
    \begin{equation}
        L_{-1}\ket{\phi_{\bar{\beta}}}=L_{1}\ket{\psi_{\bar{\beta}}}=0\,.
    \end{equation}
Suppose now that \eqref{ChodThor2} is exact. Then there exists $S\in\mathrm{End}(\mathcal{H})$ such that
\begin{equation}
    [L_{m},S]=\beta_{2}^{\alpha\bar{\alpha}}\,m\,J_{\alpha(m)}\bar{J}_{\bar{\alpha}(0)}\,.
\end{equation}
Then, for $m=-1$ we obtain
\begin{equation}
    [L_{-1},S]\ket{\phi_{\bar{\beta}}}=-\ket{\psi_{\bar{\beta}}}\,.
\end{equation}
Finally, using that $L_{-1}^{\dagger}=L_{1}$, we arrive at a contradiction:
\begin{equation}
    \braket{\psi_{\bar{\beta}}|\psi_{\bar{\beta}}}=-\bra{\psi_{\bar{\beta}}}[L_{-1},S]\ket{\phi_{\bar{\beta}}}=\bra{\psi_{\bar{\beta}}}SL_{-1}\ket{\phi_{\bar{\beta}}}-\bra{\psi_{\bar{\beta}}}L_{-1}S\ket{\phi_{\bar{\beta}}}=0\,.
\end{equation}
\end{pr}

\begin{cor}
    Under the assumptions of \textbf{Proposition 6}, if the Cardy matrix $\beta_{2}^{\gamma\bar{\gamma}}(g)\neq 0$, then the cocycle $\beta_{2}$ is not exact.
\end{cor}

\begin{rem}
It is also worth noting that the terms containing $\eta_{\alpha\beta}$ and $\bar{\eta}_{\bar{\alpha}\bar{\beta}}$ turn out to be exact and do not contribute to the cohomology class of the obstruction. Much like their vanishing contributions to the OPE structure constants in the marginal sector, they consequently fail to affect the leading order of the quantum-field-theoretic beta function.
\end{rem}
\section{Conclusions and Discussion}
In this paper, we have established a rigorous algebraic framework for $J\bar{J}$-deformations of two-dimensional conformal field theories by explicitly constructing the deformed conformal algebra module structure. Furthermore, we demonstrated that the field-theoretic beta function at the leading order is naturally recovered from the Chevalley-Eilenberg cocycle governing the deformation obstructions.

A natural next step is the construction of higher-order coefficients of the beta function as \emph{higher cohomological obstructions} to the deformation of the $\mathfrak{Conf}_{2}$ action on the space of states $\mathcal{H}$. To achieve this, one needs to explicitly construct a \emph{homotopy} $h$ for the operator $\{\mathrm{Q},\cdot\}$ or, equivalently, an explicit decomposition of the Chevalley-Eilenberg complex into its cohomology and an acyclic subcomplex. If we view the obstructions as maps 
\begin{equation}
\mu_{n}: (H^{1}_{\{\mathrm{Q},\cdot\}})^{\otimes n}\rightarrow H^{2}_{\{\mathrm{Q},\cdot\}}\,,
\end{equation}
the higher obstructions are constructed iteratively by means of the homotopy operator. Namely, the second obstruction is simply given by $\mu_{2}=\{\cdot,\cdot\}$, and the matrix element of this map evaluated for the cocycles $\delta^{(1)}\mathrm{Q}_{\alpha\bar{\alpha}}$ has been computed in \textbf{Section 2.3}. Then, for example, the third obstruction is given by the formula
\begin{equation}
\mu_{3}=\{\cdot, h(\{\cdot,\cdot\})\}\,.
\end{equation}
Based on preliminary calculations, we expect that such a homotopy can be chosen in a way that allows the cohomological subcomplex to contain the cocycle $\beta_{2}$.

Finally, we anticipate a deep relationship between the cocycles \eqref{curcur_cocycle} and \eqref{obs_cocycle}, reminiscent of the structures arising in BRST-symmetric theories \cite{Lian:1992mn}. Namely, we expect that there exists a canonical one-to-one correspondence between marginal $1$-cocycles and a certain class of $2$-cocycles, which realizes the obstructions. Such a connection would allow one to interpret the $L_{\infty}$-\emph{identities} as quadratic relations on the coefficients of the beta function.

\section*{Data availability statement}
Data availability is not applicable to this article as no data were created or analyzed during this study. All theoretical results and relevant calculations are included within the published text.

\section*{Conflict of interest statement}
On behalf of all authors, the corresponding author states that there is no conflict of interest.

\section*{Acknowledgements}
We are grateful to Albert Schwarz, Nicolai Reshetikhin and Vyacheslav Lysov for helpful discussions. Maxim Gritskov is supported by the Ministry of Science and Higher Education of the Russian Federation (Agreement No. 075-15-2025-013). Oleksandr Gamayun acknowledges support from Cognia AI.

\bibliographystyle{plain} 
\bibliography{refs}



\end{document}